# Rust: The Programming Language for Safety and Performance

William Bugden[1] and Ayman Alahmar[*,1]

[*]Corresponding Author: aalahmar@lakeheadu.ca, ORCID: **0000-0003-4011-1023**
[1]Department of Software Engineering, Lakehead University, Thunder Bay, Ontario, Canada P7B 5E1

**Abstract:** Rust is a young programming language gaining increased attention from software developers since it was introduced to the world by Mozilla in 2010. In this study, we attempt to answer several research questions. Does Rust deserve such increased attention? What is there in Rust that is attracting programmers to this new language? Safety and performance were among the very first promises of Rust, as was claimed by its early developers. Is Rust a safe language with high performance? Have these claims been achieved? To answer these questions, we surveyed and analyzed recent research on Rust and research that benchmarks Rust with other available prominent programming languages. The results show that Rust deserves the increased interest by programmers, and recent experimental results in benchmarking research show Rust's overall superiority over other well-established languages in terms of performance, safety, and security. Even though this study was not comprehensive (and more work must be done in this area), it informs the programming and research communities on the promising features of Rust as the language of choice for the future.

## INTRODUCTION

Rust began in 2006 as a personal side project of Graydon Hoare, an employee at Mozilla. Hoare explained that Rust got its name from rust fungi and as a subsequence of "robust" [1]. Mozilla saw the potential of the new language and began sponsoring the project in 2010 [1][2]. In short order, the first pre-alpha version of the Rust compiler was released in January 2012. "Despite its relative youth, Rust has steadily risen in the ranks of popular programming languages. In fact, while it ranked 33 in July 2019, by July 2020 it had risen to the 18th spot on the TIOBE Programming Community Index. Similarly, according to the Stack Overflow Developer Survey, Rust has been the "most loved" language since 2016" [2].

Rust is a systems programming language meant to supersede languages like C++. The main focus of Rust is (memory) safety, but it later began to target performance as well, adopting the C++ approach of zero cost abstraction. While outdoing other systems languages in safety and performance is great, Rust has other advantages that make it much more versatile than them. Rust has an ecosystem that greatly simplifies any software project; a number of crates (Rust libraries) exist for any need you may have, and all can be installed using the official Cargo tool.

**RELATED WORK**

Rust has been evaluated by many recent research papers in the fields of software development and programming languages. Jung et al. [3] state that there is a tension in programming language design between the following two seemingly irreconcilable desiderata: control and safety. The authors show that Rust is the first industry-supported computer programming language to overcome the longstanding trade-off between the control over resource management provided by lower-level languages for systems programming, and the safety guarantees of higher-level languages [3]. "Rust tackles this challenge using a strong type system based on the ideas of ownership and borrowing, which statically prohibits the mutation of shared state. This approach enables many common systems programming pitfalls to be detected at compile time." [3]. In a PhD Dissertation by Ralf Jung [4], the author describes Rust as a young programming language that was successful in filling the gap between "high-level languages—which provide strong static guarantees like memory and thread safety—and low-level languages—which give the programmer fine-grained control over data layout and memory management" [4]. Bugden and Alahmar [5] have conducted a recent analysis and benchmarking study among six prominent programming languages (C, C++, Go, Java, Python, and Rust), with emphasis on safety and performance. From the performance side, the authors showed that Rust surpasses Go, Java, and Python; and can keep up compared to C and C++. From the safety perspective, the authors found that Rust overall "was the most safe, especially in concurrent environments where Rust's data race prevention can avert many software bugs and vulnerabilities" [5]. The authors concluded that considering just performance, Rust is one of the best languages, while when safety is also considered, Rust is the definitive best [5].

Jung et al. [6] described a formal version of the Rust type system in order to partially prove the language's safety in regards to its ownership discipline when working with `unsafe` code. Using this formal definition the authors were able to discover a data race that could occur using a `MutexGuard` from the standard library (this is due to unsafe behavior inside an `unsafe` block, the Rust guarantees that prevent data races still hold). Additionally, the authors were able to determine that "various important Rust libraries" have safely encapsulated unsafe code, allowing developers to use these libraries with confidence in their interface being safe. The authors hope to extend this project and improve its accuracy in the future, bringing it closer to the actual language and making Rust even safer as a consequence.

Uzlu et al. [7] discussed the benefits of Rust in the context of the internet of things (IoT) space. The authors created an excellent figure that compared Rust, C, C++, Haskell, Lisp, Python, Go, and Swift on various features applicable to the IoT space like interoperability with other systems languages, various safety features, and the ability to operate without an operating system. After analyzing the safety features and specific security vulnerabilities of the past for Rust and the other languages, the authors concluded that Rust is the ideal language for IoT (and embedded systems in general) as its safety not only helps ensure security but also maximizes stability.

**RUST SAFETY**

The safety of a programming language comes down to its ability to prevent or detect errors like buffer overreads. Most of these issues come down to memory management [5], so modern languages invented memory management systems—the more well known system being garbage collection. Rust uses an ownership based system to determine memory allocations and deallocations at compile time, improving run time performance. To put it simply, memory is allocated when a variable is declared and deallocated once the variable is no longer in scope as the variable's scope is the *owner* of the memory (see Listing 1). This aspect of the ownership system prevents use after free and double free vulnerabilities by not having the programmer manually manage the memory. The full complexity of the ownership system is beyond the scope of this paper, but it should be noted that it is possible to associate memory with a lifetime that allows it to outlive its variable's scope.

The other really amazing aspect of the ownership system is that there can be only one owner of memory at a time. This leads to a problem with sharing values between functions, structs, and threads. This is solved using a borrowing system with its own rules for shared references. The rules of the borrowing system are fairly simple: there can either be one mutable reference to memory *or* multiple immutable references, meaning that there will only ever be, at most, one thread modifying the memory with no others reading. The combination of these systems and their rules completely prevent data races from occurring, making Rust an amazing choice for highly concurrent software systems.

Besides these systems, Rust has other safety features like automatic bounds checking for buffer accesses. Rust does not allow pointer arithmetic outside of the rarely needed unsafe block, so buffers are accessed via bounds checked indexes, causing a panic when an out of bounds occurs (instead of continuing in an undefined state). This prevents both buffer overreads and overflows.

```rust
fn make_vec()
{
    // owned by make_vec's scope
    let mut vec = Vec::new();

    vec.push(0);
    vec.push(1);
    // scope ends, `vec` is destroyed
}
```

Listing 1. Simple example of Rust's ownership system managing memory (from the official Rust blog [8]).

**RUST PERFORMANCE**

As mentioned in the introduction, Rust utilizes the concept of zero cost abstraction to simplify the language without limiting performance. A simple example of this idea is monomorphization, which allows one to create generic functions that are converted into the needed concrete type functions at compile time, so no runtime costs are incurred. Additionally, this concept is applied to the standard library to ensure there is no need to recreate common types like collections, providing greater performance and better interoperability of libraries [9].

The next significant benefit of Rust is its lack of a garbage collector, which has many benefits for performance. Garbage collection adds overhead at runtime as the garbage collector must track the memory in some way (usually through reference counting) in order to determine when memory can be freed. This increases both memory usage (very important for embedded devices with limited memory) and CPU usage (important for performance critical software). Another considerable problem with garbage collection is that it is more difficult to control, leading to unexpected pauses in execution when it is running and/or freeing unused resources.

The figures below are based on data from [5] and reflect the experimental performance results of different benchmarks for six languages, including Rust, for two algorithms. The Monte Carlo Pi estimation relied on a random number generator (RNG), so the third test used a standard RNG implemented for each language for a more accurate comparison than the second test which used the language's standard RNG. As can be seen from the results, Rust is the leader for two of the three algorithms in CPU time, and only loses by a small margin for the third test to C and C++; when it comes to memory usage, Rust is second only to C for all tests.

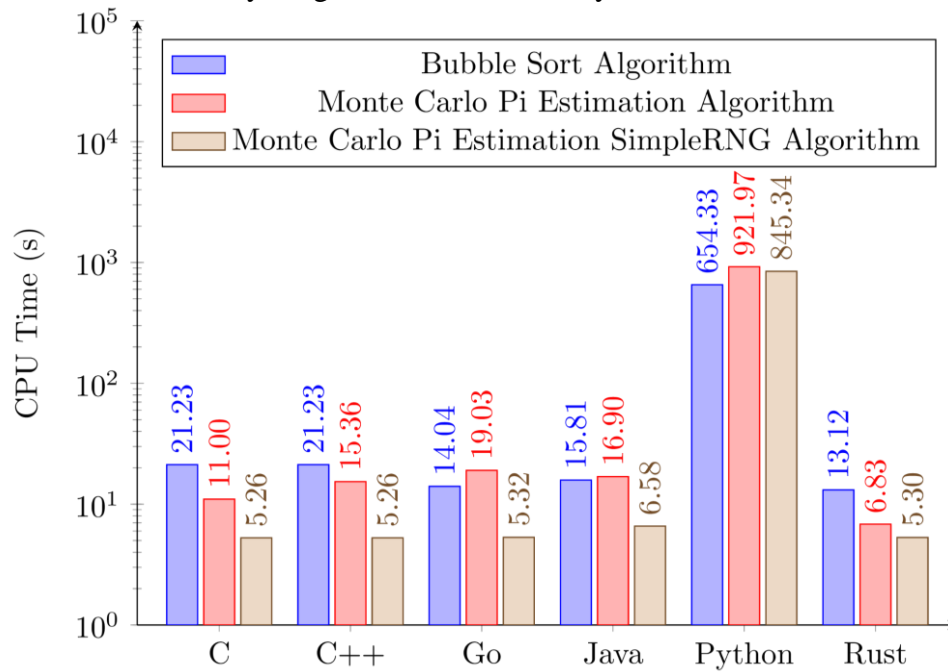

Fig 1. Average CPU time benchmark results [5].

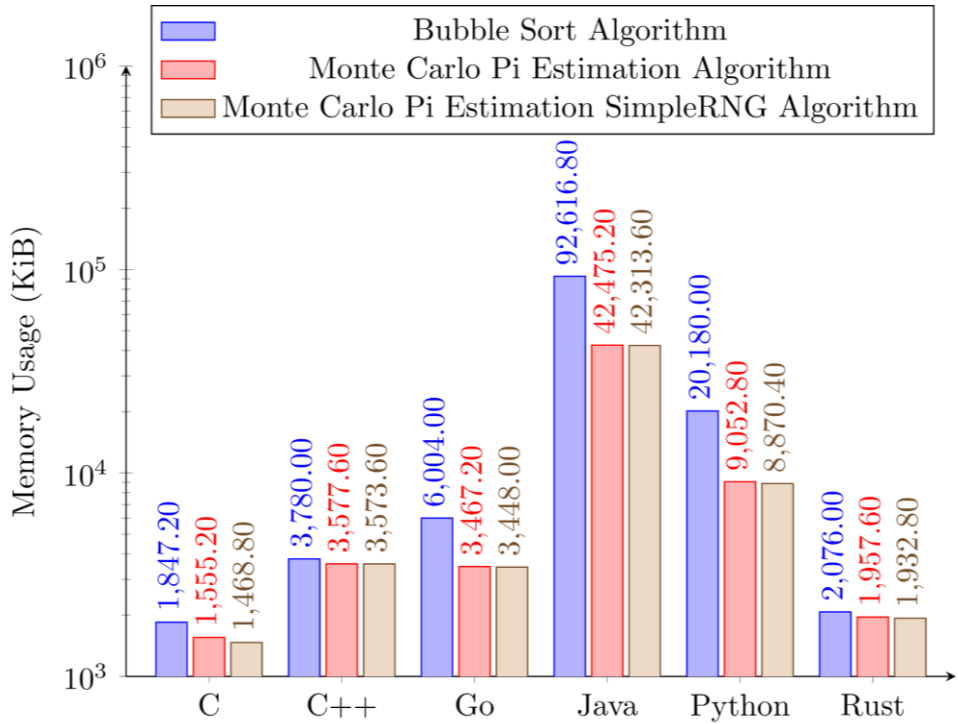

Fig 2. Average memory usage benchmark results [5].

**RUST SECURITY**

There is a lot of overlap between the safety of a language and the security of software written in that language. The Microsoft Security Response Center estimates that 70% of bugs they assign a common vulnerabilities and exposure (CVE) tag to are caused by memory exploits (i.e., buffer overflows or other memory misuses) [5]; this, obviously, means that ensuring memory safety can eliminate up to approximately 70% of security bugs.

Memory safety failures represent a large category of different bugs. We have covered some of the more common memory related bugs in previous work [5]; they include: use after free, dereferencing a null pointer, double free, buffer overflow, buffer overread, and data races (not all consider data races to be a memory safety issue). The main problem with any safety related bug is that they all result in undefined behavior (UB), which means that anything can happen, including security vulnerabilities.

Use after free bugs are most common in languages with manual memory management like C. To create one, all one must do is free a needed memory location while it is still needed. Accessing memory after being freed can cause many different things to happen, and they will be dependent on many factors such as operating system and hardware. A potential security vulnerability this could cause would be to leak information by sending the data at the freed memory location as once the memory is freed it may be used for other variables [5]. Although, if the operating system has decided the address is invalid once freed, the program will have a segmentation fault and execution will be halted. Rust does not allow manual memory management outside of the rarely needed `unsafe` block, so it does not have this problem in general and it is much easier to ensure correctness of a small section of code than for the entirety of the codebase.

Null pointer dereferencing occurs when attempting to dereference a null pointer (generally a pointer pointing to address 0). This problem has been called "the billion dollar mistake" [10] as it generally causes systems to be left in an undefined state rather than causing an outright failure, leading to any number of imaginable problems. Consider for a moment a program that attempts to copy data from one memory location to another, but the source is a null pointer. What is copied (if anything at all) could be anything—maybe it is just all zeroes or maybe it is random bits from address 0—but it will have side effects that spread through the rest of the execution, potentially causing security vulnerabilities and random behavior that is difficult to diagnose. Rust solves this problem by not having null values at all, instead using a type called `Option` to represent possibly uninitialized values. This ensures safety and security, and improves the ergonomics of the language because it is often necessary to declare a variable before initialization.

Double frees can lead to similar problems as use after frees as it is possible the address attempting to be freed a second time is now in use for another variable. The other scenario is the address is not being reused and the behavior is undefined. As mentioned previously UB can produce security vulnerabilities, but it is difficult to predict how as all UB is dependent on the specific compiler, operating system, and hardware.

Buffer overflows are a well known exploit as C, and its standard library, have created an abundance of vulnerabilities. Buffer overflows can allow an attacker to insert their own instructions to be executed, potentially leading to a complete takeover of the system without much difficulty once discovered. Alternatively, an attacker can modify variables in memory that may give them access to higher privileges. Preventing buffer related exploits is trivial using a bounds check, which many languages do automatically at runtime. Rust bounds checks all array access [9], although they may be, and often are, safely elided by the LLVM back end compiler for performance.

Buffer overreads are in a similar vein as overflows, but are generally less of an issue because they only leak information rather than allow for a full takeover (unless the information can be used to take over the system, like a password). This can still have disastrous consequences as in the case of the Heartbleed vulnerability that affected an estimated 24-55% of the Alexa top one million HTTPS-enabled websites at the time of disclosure [9]. All of these sites were vulnerable to having potentially sensitive data read from the memory of their server. Rust does not allow buffer overreads and insteads panics (see Listing 2) when they occur thanks to its automatic buffer bounds checking [9].

Data races are unpredictable by their nature, causing different results each time they occur. This makes them difficult to debug and can lead to exploits based on timing. For example, picture a server running software that handles requests of some kind (like a web application API back end); this software has three threads, one handles the actual communication and the other two do some kind of processing for the request. It could be possible for an attacker to send two requests with both requests attempting to write to the same memory, leading to a potential data race. Depending on what the server actually does this could allow several different exploits, like data corruption, leaking of data, or some type of privilege escalation. Few languages enforce safety for concurrent data access, although many offer atomic types that prevent data races using locks (although this can reduce performance). Rust prevents data races using its ownership and borrowing rules, which only allow one thread to read or write data, or multiple threads to read but not write data at a time [5][9].

```rust
fn main()
{
    let buffer = [34, 22, 75, 54, 88];

    for i in 0..6
    {
        println!("Value at index {} is {}", i, buffer[i]);
    }
}
```

Listing 2. Rust buffer overread example that will cause a
panic at runtime rather than expose data in memory.

**OTHER FEATURES OF RUST**

Rust offers many modern features that the more established systems languages tend to lack. The first major tool of the Rust ecosystem is Cargo, which is a build system and package manager [9][12]. Cargo facilitates almost anything you may need in a software project, including dependency management (downloading and compiling), build management (debug, release, and custom profiles), testing (unit testing and integration testing), benchmarking of performance, documentation generation, and can install additional tools that integrate with it.

Dependency management is handled via a configuration file that includes project dependencies; the dependencies are then automatically installed during compilation. Finding crates (libraries) to use for your project is simple using the official Rust community crate registry (hosted at crates.io).

Building, testing, benchmarking, and documentation generation are all built in features of Cargo that work out of the box [12]. Builds can be configured in the project's configuration file (Cargo.toml) with different profiles for any need you may have. By default there is a developer build and release build that maximize certain aspects of the compiler appropriate for those scenarios (developer builds are fast to build and include debug symbols; release builds are slower to build but faster to execute). Cargo can perform unit testing and integration testing using its test command. Unit tests are written with the code they test and will only be compiled when testing; integration tests are written in their own files in the tests directory. Benchmarking is done similarly to unit testing with benchmarking code in the same file as the function being benchmarked, and only being compiled during benchmarking [9][12]. Finally, there is documentation generation which is done from special comments in the code. Documentation is generated as a static HTML website for easy viewing in web browsers and can be hosted online.

One great tool to install with cargo is Clippy, which is a linter that provides extra warnings and can enforce extra compile time constraints, making code even safer and faster. It is highly recommended in the Rust community with many projects using it to ensure code quality of pull requests [11].

Another important tool that can be installed with Cargo is rustfmt, which is a code formatter for Rust that offers extensive configuration options. This is even more of a must for any source controlled projects as it ensures code formatting consistency. You could even create your own configuration for your local copy of a project and format the code before committing, finally ending the spaces vs. tabs war.

These are just some of the tools available in the Rust ecosystem, there are many more—Cargo itself has many more features explained in full in its documentation [12]. This tooling alone makes Rust a much better development experience than most systems languages and is most likely a considerable contributor to its rise in popularity. Additionally, Rust is approaching the status of a mainstream language in health informatics applications, a field that was previously dominated by other languages [13-17].

Rust has made steady progress in capturing programmers' interest over its years of existence. Fig. 3 below shows the worldwide increase in interest in Rust among the software development community for the past ten years (May 2012 - May 2022).

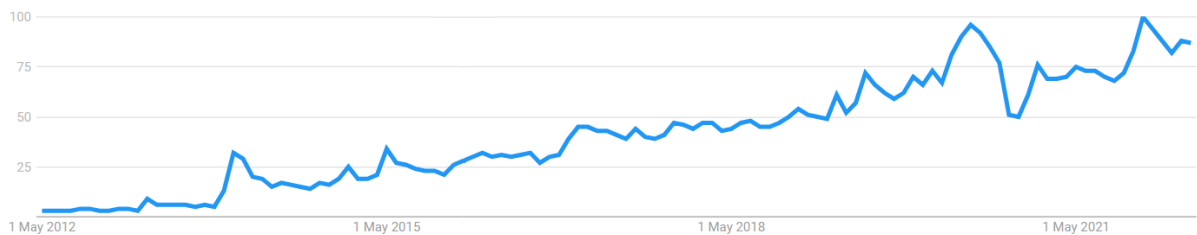

Fig 3. Interest in Rust programming language over time
(May 2012-May 2022, generated from google trends) [18].

**CONCLUSION**

According to the Stack Overflow Developer Survey, Rust has been the most loved language for six years in a row (since 2016). This study showed that this "love" is justified. Since its inception, Rust was designed for performance and safety, especially safe concurrency. Rust is outstanding for enforcing memory safety where all references point to valid memory without requiring the use of a garbage collector. Through recent research on Rust, this new language was found to outperform well-established languages in terms of safety. Rust was found to be the safest language compared to C, C++, Java, Go, and Python. Performance-wise, Rust was also among the top, which makes it remarkable from the point of view of one language performing well in both safety and performance (note that the trend before Rust was to compromise safety for performance or vice versa). Other positive features of Rust include its sizable ecosystem of tooling and libraries; not many languages offer anywhere near as many standard and well supported community tools. Having said that about Rust, it is worth mentioning that Rust has received some criticism, and not all studies on Rust have been partial to the language. The criticisms of Rust tend to originate from its lack of maturity. Rust is only 12 years old compared to C being 50 years old. Also, Rust needs more adoption to become a mainstream language. For example, C and C++ are well-adopted and much more established in the industry than Rust, making it easier to find and employ developers in these languages. In contrast, there is a lack of demand for Rust developers in the market. In addition, a limitation of this study is that it is not a comprehensive study; therefore, we recommend more collaborative work in this domain to better inform the software developer community. Nevertheless, despite the potential drawbacks mentioned above, the Rust community is growing fast, and we conclude that Rust is a highly promising language for the future generation of safe, secure, and well-performing software applications.